\documentclass[conference]{IEEEtran}
\IEEEoverridecommandlockouts
\usepackage{cite}
\usepackage{amsmath,amssymb,amsfonts}
\usepackage{algorithmic}
\usepackage{graphicx}
\usepackage{fancyvrb}
\usepackage{hyperref}
\usepackage{textcomp}
\usepackage{xcolor}
\def\BibTeX{{\rm B\kern-.05em{\sc i\kern-.025em b}\kern-.08em
    T\kern-.1667em\lower.7ex\hbox{E}\kern-.125emX}}

\usepackage{cleveref}

\usepackage{flushend}

\makeatletter
    \newcommand{\linebreakand}{%
      \end{@IEEEauthorhalign}
      \hfill\mbox{}\par
      \mbox{}\hfill\begin{@IEEEauthorhalign}
    }
\makeatother

\begin{document}

\title{Smart Contracts, Smarter Payments: Innovating Cross Border Payments and Reporting Transactions}

\author{\IEEEauthorblockN{Maruf Ahmed Mridul}
    \IEEEauthorblockA{\textit{Department of Computer Science} \\
    \textit{Rensselaer Polytechnic Institute}\\
    Troy, NY, USA \\
    mridum@rpi.edu}
\and
\IEEEauthorblockN{Kaiyang Chang}
    \IEEEauthorblockA{\textit{Department of Computer Science} \\
    \textit{Rensselaer Polytechnic Institute}\\
    Troy, NY, USA \\
    changk2@rpi.edu}
 \linebreakand
\IEEEauthorblockN{Aparna Gupta}
 \IEEEauthorblockA{\textit{Lally School of Management} \\
 \textit{Rensselaer Polytechnic Institute}\\
 Troy, NY \\
 guptaa@rpi.edu}
\and
\IEEEauthorblockN{Oshani Seneviratne}
    \IEEEauthorblockA{\textit{Department of Computer Science} \\
    \textit{Rensselaer Polytechnic Institute}\\
    Troy, NY, USA \\
    senevo@rpi.edu}
}

\maketitle

\begin{abstract}
The global financial landscape is experiencing significant transformation driven by technological advancements and evolving market dynamics. Moreover, blockchain technology has become a pivotal platform with widespread applications, especially in finance. Cross-border payments have emerged as a key area of interest, with blockchain offering inherent benefits such as enhanced security, transparency, and efficiency compared to traditional banking systems. This paper presents a novel framework leveraging blockchain technology and smart contracts to emulate cross-border payments, ensuring interoperability and compliance with international standards such as ISO20022. Key contributions of this paper include a novel prototype framework for implementing smart contracts and web clients for streamlined transactions and a mechanism to translate ISO20022 standard messages. Our framework can provide a practical solution for secure, efficient, and transparent cross-border transactions, contributing to the ongoing evolution of global finance and the emerging landscape of decentralized finance.
\end{abstract}

\begin{IEEEkeywords}
Blockchain, Smart Contracts, Cross-Border Payments, ISO20022, FinTech, Interoperability, Decentralized Finance.
\end{IEEEkeywords}

\section{Introduction}

Cross-border payments face several significant challenges affecting efficiency, cost, and accessibility. Traditional international transactions are often slow, taking days to complete due to the involvement of multiple intermediaries such as correspondent banks~\cite{he2021digitalization}. Each intermediary adds time and increases transaction fees, making cross-border payments expensive. Moreover, the lack of transparency in the process can lead to uncertainties regarding transaction status and final settlement times. Regulatory compliance is another hurdle, as transactions must adhere to the varying financial regulations of each country involved, which includes fulfilling anti-money laundering (AML) and combating the financing of terrorism (CFT) requirements. Additionally, currency conversion in these transactions introduces further complexity and cost, with fluctuating exchange rates potentially affecting the final amounts received by beneficiaries. These issues create a cumbersome, opaque, and costly process that impacts consumers, businesses, and financial institutions engaged in global trade and remittances.

The quest for a solution to this problem has been ongoing throughout international commerce, where the need to facilitate payments across borders has always been critical. In response to these enduring challenges, emerging fintech solutions, particularly those involving Decentralized Finance (DeFi) using blockchain technology, could provide promising avenues that could potentially revolutionize cross-border payments within the next decade~\cite{bindseil2022towards}.
Blockchain’s inherent attributes—decentralization, immutability, and transparency—offer a new paradigm for addressing the inefficiencies and complexities of traditional payment systems.

Recognizing the transformative potential of blockchain technology, we contribute to advancing cross-border payment systems by developing a framework that leverages blockchain technology and smart contracts to enable seamless, secure, and efficient international transactions. Our contributions include:

\begin{itemize}
    \item Proposing a novel framework for cross-border payments leveraging blockchain technology in an interoperable ecosystem.
    \item Implementation of the proposed framework, integrating smart contracts and web clients to streamline transactions and ensure transparency, efficiency, and security.
    \item Development of mechanisms for translating ISO20022 standard messages using a web client, facilitating seamless communication and interoperability across disparate systems. 
\end{itemize}

\section{Background}


\subsection{Blockchain Technologies}

Over the past decade, technological advancements have reshaped the landscape of these transactions, with blockchain technology emerging as a disruptive force. One of the notable shifts has been the growing prominence of blockchain technology in facilitating cross-border payments \cite{deng2020application}. Blockchain, originally introduced as the underlying technology behind cryptocurrencies like Bitcoin \cite{crosby2016blockchain}, has evolved into a versatile platform with applications across various sectors, including finance.
The decentralized and immutable nature of blockchain offers inherent advantages for cross-border payments, promising enhanced security, transparency, and efficiency compared to traditional banking systems. As a result, academia and industry have turned their attention to exploring blockchain's potential to revolutionize international financial transactions.

Originally defined by Nick Szabo~\cite{szabo1996smart} and popularized by Vitalik Buterin on the Ethereum ecosystem~\cite{buterin2014next}, smart contracts are self-executing code that is deployed across a distributed, decentralized blockchain network, which results in transactions that are trackable and irreversible. 
Since smart contracts automate and enforce the terms of agreements without intermediaries, they offer potential cost savings and efficiency gains in international transactions \cite{vinayak2019design}. Meanwhile, DeFi protocols leverage blockchain technology to create decentralized financial ecosystems that offer open, interoperable, and transparent financial services \cite{schar2021decentralized}.
However, there is a critical interoperability issue among different blockchain platforms, with several blockchain interoperability platforms developed to address the challenges and risks associated with adopting a single blockchain implementation~\cite{kang2022blockchain}. In our CBPR+ implementation, we assumed EVM-compatible interoperable blockchains would interact seamlessly to ensure compliance with international standards such as ISO20022. 

\subsection{Cross-Border Payments and Reporting Plus (CBPR+)}

CBPR+ is an industry-led initiative that aims to standardize global payment transactions and harmonize standards. It focuses on improving the efficiency of cross-border payments. 
It utilizes the ISO20022 \cite{mcgill2023iso} messaging format to ensure a consistent customer experience across geographical boundaries. ISO20022 offers a richer and more flexible standard compared to its predecessors, facilitating innovation and efficiency in cross-border payment systems.
Adopting international standards, such as ISO20022, is essential in standardizing communication protocols and data formats in financial transactions \cite{bouille2019adoption}.
The adoption and impact of ISO20022 in facilitating effective communication and standardization in financial transactions are critically examined by Bouille and Haase \cite{bouille2019adoption} and further elaborated by McGill \cite{mcgill2023iso}. Their discussions emphasize the importance of this standard in ensuring interoperability and efficiency, which is crucial for the framework our research seeks to develop.

The CBPR+ process involves a series of ISO20022 message exchanges facilitating seamless fund transfers across borders. Key messages include but are not limited to the following:

\begin{itemize}
    \item \texttt{pacs.008}: Customer Credit Transfer Initiation message
    \item \texttt{pacs.002}: Payment Status Report message
    \item \texttt{pacs.004}: Payment Return message
\end{itemize}

The term \texttt{pacs.008} is a specific identifier within the ISO 20022 standard for financial messaging, defining various messages for financial transactions. The abbreviation \texttt{pacs} stands for ``Payments Clearing and Settlement," and the \texttt{.008} identifies the specific message type for customer credit transfers. 
These messages form the foundation of transaction management within CBPR+, covering initiation, status reporting, and error handling.

\subsection{Blockchain for Cross Border Payments}

Emerging technologies like smart contracts and DeFi present new opportunities for streamlining cross-border payments. 
However, despite the promise of blockchain and related technologies, challenges persist in achieving widespread adoption and scalability in cross-border payments.
Many issues, such as transaction finality, regulatory compliance, and user experience, remain areas of active research and development \cite{zilnieks2023cross, he2021digitalization}. 
In exploring blockchain technology for cross-border payments, blockchain interoperability emerges as a crucial element, profoundly discussed by Mohanty et al. \cite{mohanty2022blockchain} and Belchior et al. \cite{belchior2021survey}. These studies underscore the necessity of seamless integration across different blockchain systems to enable more robust and sustainable financial infrastructures, which directly informs the objectives of our project.

Deng \cite{deng2020application} analyzes the application modes and advantages of blockchain technology in cross-border payments, highlighting its potential to address the shortcomings of traditional cross-border trade. Deng's insights provide valuable context for understanding the challenges and opportunities of leveraging blockchain for international transactions. 
Similarly, Isaksen \cite{isaksen2018blockchain} investigates how blockchain technology may benefit banks' positions in the cross-border payment segment. By exploring the potential benefits and challenges of utilizing blockchain in financial institutions, Isaksen offers insights into the practical implications and adoption barriers of blockchain technology in real-world financial systems. This understanding is crucial for designing solutions that align with the needs and constraints of existing financial infrastructures.
He \cite{he2021digitalization} analyzes the broader context of digitization in cross-border payments, noting significant global efforts to enhance transactional efficiency. This perspective is valuable for understanding the environmental and systemic changes influencing our research direction. Meanwhile, Le Quoc et al. \cite{le2022credit} offer a specific example of blockchain application in the letter-of-credit system for international trade, presenting a case study that mirrors the potential applications of our research.
Zilnieks and Erins \cite{zilnieks2023cross} explore cross-chain bridges to standardize distributed ledger technologies, a discussion particularly relevant to our project's aim to enable smooth interoperable transactions across different blockchain platforms. Another related work by Schär \cite{schar2021decentralized} discusses DeFi, providing insights into the infrastructure and potential of blockchain-based financial markets to offer more open, interoperable, and transparent services.
Together, these contributions provide a comprehensive backdrop against which our research is situated. They illuminate the current state of blockchain technology in financial applications and highlight the diverse approaches and potential enhancements that our project could integrate. By leveraging these insights, our framework aims to contribute effectively to the evolving landscape of global financial transactions, offering solutions that are not only technologically advanced but also aligned with global financial practices and standards.

Our proposed solution aims to enable seamless, secure, and efficient international transactions leveraging blockchain technologies and existing standards to provide interoperability with existing systems and regulatory compliance.
At the heart of our transaction system lies the smart contracts, robust and autonomous entities deployed on Ethereum Virtual Machine (EVM) compatible blockchains to demonstrate the system's interoperability for smarter payments in cross-border scenarios. 
Our smart contract-based implementation embodies the principles of transparency, efficiency, and security within the decentralized network and is also interoperable through standards.

\section{Methodology}

Our methodology includes a blockchain-based framework to replicate and enhance the CBPR+ process. It uses smart contracts and web clients to streamline and secure cross-border transactions within a decentralized environment.
Within our framework, the smart contract serves as an autonomous agent managing the logic and state of each transaction, while the web client acts as the interface between traditional banking systems and a blockchain. Together, they emulate the CBPR+ process in a distributed ledger environment.

\subsection{Detailed System Architecture}

\begin{figure*}
    \centering
    \includegraphics[width=\linewidth]{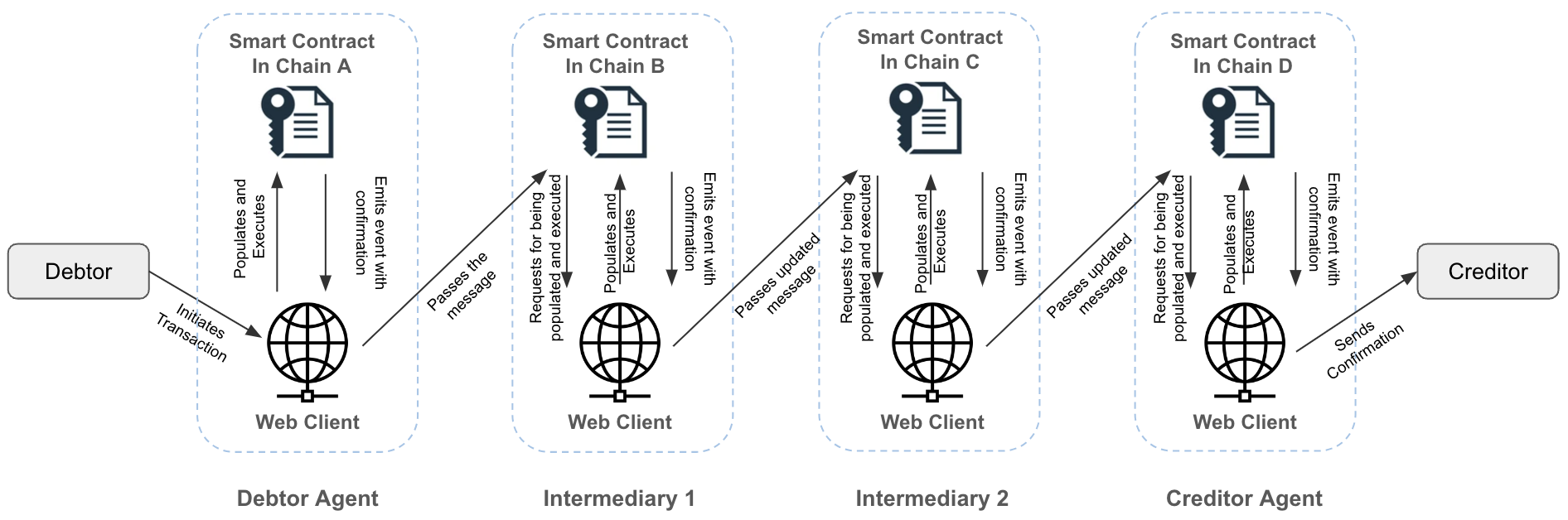}
    \caption{Workflow of the Framework for an Example Transaction involving a Debtor and Creditor for a Cross Border Payment.}
    \label{fig:impl_workflow}
\end{figure*}

To illustrate our framework's functionality, let's consider a simulated CBPR+ transaction scenario involving multiple banks across different blockchains. Refer to Figure \ref{fig:impl_workflow} for a depiction of the following discussion. The workflow initiates with the debtor, who provides transaction instructions. The debtor's agent bank (Bank A) communicates these instructions to its corresponding smart contract, which debits the debtor's account and emits a confirmation message. Subsequently, the web client generates an ISO20022 \texttt{pacs.008} message for the next agent and triggers the respective smart contract. This process iterates sequentially through intermediary banks until reaching the creditor agent, concluding the transaction.

\subsubsection{Smart Contracts}
Engineered to emulate and enhance the CBPR+ methodology, the smart contracts encapsulate the essential functions and rules governing cross-border payment transactions, from initialization and access control to transactional validation and event-driven communication.

\begin{itemize}
    \item \textbf{Initialization and Access Control:}
    This smart contract establishes an access control hierarchy upon deployment, designating the deploying address as the owner with administrative privileges. Role-based access control mechanisms ensure that only authorized entities can execute critical functions that alter the contract's state or configuration. By enforcing access restrictions, the contract safeguards against unauthorized manipulation or misuse, preserving the integrity and reliability of the transaction system.

    \item \textbf{Account Management:}
    This smart contract manages two distinct account types: \emph{Nostro Accounts} and \emph{General Accounts}. Nostro Accounts facilitate holding funds in foreign currencies, essential for cross-border transactions, while General Accounts cater to everyday transactional needs. Account creation within the contract involves specifying a unique account number and distinguishing between account types. The specific account management actions include mapping and tracking account addresses, verifying account existence and uniqueness, and maintaining a state variable for tracking available funds in each account. 

    \item \textbf{Transaction Details:}
    \emph{Solidity structs} define structured data entities to capture and process transaction details. These structs encapsulate essential transactional information, including currency and amounts (\texttt{Amount}), debtor instructions (\texttt{DbtrInstruction}), and comprehensive transaction metadata (\texttt{MsgInfo}). Mirroring the ISO20022 message structures, these defined structs enable the seamless integration of cross-border payment transactions within the blockchain framework. 
    For example, the following code snippet shows the \texttt{DbtrInstruction} struct in the smart contract. 

    \begin{Verbatim}[fontsize=\footnotesize]
    struct DbtrInstruction {
        Amount IntrBkSttlmAmt;
        address DbtrAgt;
        string DbtrAcct;
        string DbtrAgtIsoMsg;
        address NxtAgt;
    }
    \end{Verbatim}

    For this specific data structure, the web client extracts the required information from the XML-formatted \texttt{pacs.008} message using the following Python function, where the specific attributes correspond directly to the ISO20022 data schema to facilitate ease of integration and maintain consistency.
    This implementation enhances interoperability and compatibility with external systems and messaging standards by standardizing transactional data formats.

\begin{Verbatim}[fontsize=\footnotesize]
def get_debtor_instructions(xml_data):

    dict_data = convert_to_dict(xml_data)

    dbtr_instruction = {
      "IntrBkSttlmAmt": dict_data['Document']
            ['FIToFICstmrCdtTrf']['CdtTrfTxInf']
            ['IntrBkSttlmAmt'],
      "DbtrAgt": dict_data['Document']
            ['FIToFICstmrCdtTrf']['CdtTrfTxInf']
            ['DbtrAgt']['FinInstnId']['BICFI'],
      "DbtrAcct": dict_data['Document']
            ['FIToFICstmrCdtTrf']['CdtTrfTxInf']
            ['DbtrAcct']['Id']['Othr']['Id'],
      "DbtrAgtIsoMsg": xml_data,
      "NxtAgt": dict_data['Document']
            ['FIToFICstmrCdtTrf']['GrpHdr']
            ['InstdAgt']['FinInstnId']['BICFI']
    }

    return dbtr_instruction
\end{Verbatim}

    \item \textbf{Transactional Methods:}
    The key to the smart contract's functionality is transactional methods, which are responsible for initiating, validating, and processing cross-border payment transactions. The functions in the smart contract, alongside making the transactions, serve as gatekeepers, ensuring the security, integrity, and validity of fund transfers. They validate account existence, balances, and transaction parameters before initiating fund transfers. They also incorporate comprehensive checks such as control sum verification and nostro account balance validation. 

    \item \textbf{Event-Driven Confirmation Mechanism:}
    The smart contract employs an event-driven confirmation mechanism to facilitate real-time communication and transactional synchronization. At crucial transactional milestones, the contract emits events such as \texttt{MakeTransfer} and \texttt{PassISOMessageAlong}, signaling transaction state changes and facilitating communication with external systems, including web clients that correspond to different organizations within our simulated system. These events serve as immutable audit trails, providing a transparent history of transaction actions and enabling stakeholders to verify transactional integrity and compliance with regulatory standards. By leveraging event-driven mechanisms, the contract enhances transparency, accountability, and trust within the cross-border payment ecosystem.
\end{itemize}


\subsubsection{Web Client}

The web client plays an important role within our framework, bridging traditional banking systems and the decentralized blockchain network. Its primary function is to orchestrate the flow of cross-border payment transactions, ensuring seamless communication, data integrity, and transactional transparency. The web client's major roles are as follows:

\begin{itemize}
    
    \item \textbf{Message Parsing and Construction:}
    This component ensures the accurate interpretation and transmission of ISO20022 messages within our framework. The web client dissects incoming \texttt{pacs.008} messages, extracting essential transaction details such as the debtor's instructions, payment amounts, and beneficiary information.
    
    Upon parsing the incoming message, the web client structures all the transaction requests in a format compatible with the smart contract's defined data structures. This process involves mapping parsed data to corresponding fields within the \texttt{DbtrInstruction} struct, ensuring integration with the contract's transactional methods.
    
    Moreover, the web client continuously monitors contract events to facilitate the modification of ISO messages. When triggered by events such as \texttt{MakeTransfer} or \texttt{PassISOMessageAlong} emitted by the smart contract, the web client takes in the updated ISO message as an instruction set for the next agent in the process.
    
    \item \textbf{Transaction Initiation and Propagation:}
    Facilitating transaction initiation and propagation is a core function of the web client, serving as the intermediary between user instructions and smart contract execution. Upon receiving debtor instructions, the web client initiates transactions by interfacing with the \texttt{initiate\_transfer} function of the smart contract.
    
    In this process, the web client extracts pertinent details from the debtor's instructions, such as payment amounts, beneficiary information, and transaction metadata. It then passes this structured data to the smart contract, initiating the transaction. Concurrently, the web client monitors transaction progress through emitted contract events, updating transactional states and facilitating subsequent actions.
    As transaction events unfold, the web client dynamically propagates transaction updates to relevant parties within the blockchain network. 
    
    \item \textbf{Event Handling:}
    The web client's event-handling capabilities are important in maintaining transactional integrity and facilitating real-time communication between the blockchain framework and external systems. Designed to respond to contract events triggered during transaction execution, the web client employs event-driven mechanisms to synchronize transactional states and propagate updates.
    
    Upon receiving event notifications from the smart contract, the web client dynamically triggers corresponding actions based on predefined event handlers. For instance, when notified of a \texttt{MakeTransfer} event indicating the need to craft an outgoing ISO message, the web client invokes message construction routines, incorporating transactional details and metadata.
    
    Similarly, \texttt{PassISOMessageAlong} events signal the requirement to transmit updated ISO messages to subsequent transactional participants, prompting the web client to initiate message transmission protocols. 
    
    \item \textbf{Interaction with Smart Contract:}
    The web client and smart contract form our framework's backbone of transactional operations. Through seamless communication channels, the web client orchestrates transaction execution and state management, leveraging the capabilities of the underlying smart contract infrastructure.
    
    When triggered by user instructions or contract events, the web client interfaces with the smart contract through structured function calls, passing relevant transactional data and metadata. These function calls invoke predefined contract methods responsible for transaction initiation, validation, and propagation, ensuring the seamless execution of cross-border payment transactions.
    
    Furthermore, the web client monitors contract events, dynamically responding to state changes and transactional progress. Upon receiving event notifications from the smart contract, the web client initiates corresponding actions, updating transactional states and propagating transactional updates to relevant stakeholders.
    
\end{itemize}

The symbiotic relationship between the web client and the smart contract facilitates the seamless execution and management of cross-border payment transactions within the blockchain framework, ensuring transactional integrity, transparency, and efficiency.

\subsection{Security Protocols}
Our framework incorporates comprehensive security measures at various levels of the transactional lifecycle to safeguard transactional integrity and protect sensitive financial data.

\begin{itemize}
    \item \textbf{Access Control:} Our smart contract implements stringent access control mechanisms, restricting critical functions to authorized entities such as account holders or institutional owners. By validating user identities and permissions, the contract ensures that only authorized parties can execute sensitive operations, mitigating the risk of unauthorized access or manipulation.
    
    \item \textbf{Transaction Integrity:} 
    Leveraging structured data and Solidity's type system, our smart contract enforces transaction validation protocols, verifying transactional parameters and data integrity before processing. By encoding transactions within predefined data structures and enforcing type validation, the contract mitigates the risk of fraudulent or malicious transactions, safeguarding transactional integrity and consistency.
    
    \item \textbf{Audit Trails:} Our framework maintains comprehensive audit trails through event-driven mechanisms, providing transparent and verifiable transaction histories. Contract events such as \texttt{MakeTransfer} and \texttt{ PassISOMessageAlong} are recorded in immutable audit logs, capturing key transactional actions and state changes. These audit trails enable stakeholders to trace transactional activities, verify compliance with regulatory standards, and detect anomalous behavior.
\end{itemize}


\section{Evaluation}

In this section, we evaluate the performance and cost-effectiveness of our blockchain-based framework for CBPR+. We present gas consumption metrics obtained from a local development environment using the Ganache Command Line Interface\footnote{Ganache (\url{https://archive.trufflesuite.com/ganache}) is a command-line tool designed for Ethereum developers to create and manage personal Ethereum blockchains for testing and development purposes.} and the live testnet deployment on the Sepolia test Ethereum network\footnote{Sepolia (\url{https://sepolia.etherscan.io}) is one of the several test networks used by the Ethereum blockchain for development and testing purposes. Unlike the Ethereum mainnet, where real transactions occur with actual economic value, Sepolia provides a controlled environment where developers can test new applications, smart contracts, and upgrades without financial risk. }. Additionally, we discuss the implications of gas consumption on transaction fees and contract deployment costs.

\subsection{Gas Consumption Metrics}

Gas consumption metrics provide insights into the computational efficiency and resource utilization of smart contract operations within the blockchain network. Table~\ref{tab:gas_report} summarizes the gas consumption statistics obtained from our gas report analysis during testing.

\begin{table}[htbp]
    \caption{Gas Consumption Metrics (Gas Units)}
    \label{tab:gas_report}
    \centering
    \begin{tabular}{|l|l|l|l|l|l|}
        \hline
        \textbf{Function Name} & \textbf{min} & \textbf{avg} & \textbf{median} & \textbf{max} & \textbf{\# calls} \\ \hline
        create\_account & 26660 & 43201 & 46560 & 69676 & 17 \\ \hline
        deposit & 30351 & 38939 & 38939 & 47527 & 12 \\ \hline
        get\_balance & 921 & 921 & 921 & 921 & 14 \\ \hline
        initiate\_transfer & 121580 & 121580 & 126279 & 131428 & 12 \\ \hline
        make\_transfer & 135213 & 135213 & 140352 & 146192 & 12 \\ \hline
    \end{tabular}
\end{table}

The gas report provides detailed information on gas consumption for each smart contract function, including minimum, average, median, and maximum gas costs per function call. These metrics offer insights into the computational complexity and resource requirements of individual operations within the smart contract.

\subsection{Transaction Fees and Contract Deployment Costs}

Transaction fees and contract deployment costs are critical considerations for assessing the economic viability of blockchain-based solutions. The following metrics indicate the cost of executing transactions and deploying smart contracts on the \textit{Sepolia Testnet}. The deployed contract address on Sepolia is \href{https://sepolia.etherscan.io/tx/0x7e97eb320734ab481a84dbcd2cf26776c46ee55525652d4b3382f6558a2dd220}{0x7bD82fFA76A4a45ddF468c7106536354c9cc6909}.

\begin{itemize}
    \item \textbf{Transaction Fees:} 0.003543387976528597 ETH
    \item \textbf{Gas Fees:} 3.095497367 Gwei
\end{itemize}

Transaction fees are calculated based on the gas consumed by each transaction, while gas fees represent the price per unit of gas specified in Gwei\footnote{GWei, which is a commonly used unit of measurement in Ethereum. Wei is the smallest denomination of Ether, the native cryptocurrency of Ethereum. One GWei is equivalent to one billion wei.}.

\section{Discussion}

Our implementation of the blockchain-based framework for cross-border payments, leveraging smart contracts and web clients, represents a significant step towards enhancing the efficiency and security of global financial transactions. By replicating and streamlining the CBPR+ process within a decentralized environment, our framework offers a promising solution to the challenges associated with traditional cross-border payment systems.

\subsection{Flexibility and Customizability}

One key advantage of our framework is its flexibility and customizability, which allows it to accommodate diverse transactional requirements and be robust against any changes to regulatory frameworks governing smart contract applications. 
Financial institutions deploying our framework have the flexibility to customize the smart contract logic and message parsing rules according to their specific needs and compliance requirements. 
As long as the financial institutions can determine which portions of the ISO20022 messages are necessary to process their existing transactions, they can easily modify the extracted message fields and implement the smart contract logic within our framework to achieve their specific use cases. Additionally, financial institutions can ensure compliance with various regulatory requirements by embedding specific rules and standards directly into the smart contracts. 
This customization flexibility ensures the framework remains agile and responsive to evolving market demands and regulatory changes.
It is important to note that legacy systems must be made blockchain-compatible to integrate with our framework. 
To facilitate smooth integration with existing legacy systems in financial institutions, a future version of our framework could include Web3-enabled adapters or middleware solutions that bridge the gap between the legacy infrastructure and our smart contract-based framework, ensuring seamless data flow and operational continuity.

\subsection{Smart Contract Logic}

The smart contract is the backbone of our transaction system, encapsulating the core functions and rules governing cross-border payments. However, the logic embedded within the smart contract is scenario-dependent and may vary based on the type of transaction and regulatory framework. Our implementation provides a starting point for financial institutions to build upon, allowing them to update and modify the smart contract logic to align with their unique business processes, use cases, and regulatory obligations. 
However, we must consider certain factors when implementing smart contracts across various blockchain platforms concerning their termination mechanisms~\cite{seneviratne2024feasibility}, especially in light of regulatory changes such as the EU Data Act.
We may also need to handle unexpected situations that may arise during normal business operations relating to cross-border payment activities, such as account recovery in case of a catastrophic failure in one of the entities engaged in the trade~\cite{zhu2019proposal}. In such cases, we could leverage supplementary techniques to strengthen the smart contract logic, such as computational social choice mechanisms~\cite{liu2019strengthening} and ontology-aided mechanisms~\cite{mohsin2019ontology}.
This flexibility empowers financial institutions to adapt the framework to their specific use cases while maintaining compliance with relevant regulations.

\subsection{Message Parsing and Information Extraction}

Similarly, the web client's predefined instructions for message parsing and information extraction represent a starting point for processing ISO20022 messages within the blockchain framework. However, the specific information required for transaction processing may vary depending on the transaction type and regulatory requirements. Financial institutions deploying our framework must assess their specific information needs and customize the message parsing rules accordingly. Providing an easily configurable and extensible framework empowers financial institutions to tailor the system to their unique operational requirements while ensuring interoperability and compliance with industry standards.

Drawing parallels from our prior work, the BlockIoT project introduced a robust mechanism for integrating data from various sources in an interoperable manner that allows seamless incorporation of diverse data types, ensuring comprehensive data availability for subsequent processing in healthcare scenarios~\cite{shukla2021blockiot}. Similarly, in our financial framework, integrating various data sources—such as customer information, transaction history, and regulatory guidelines—can enhance the completeness and accuracy of cross-border payment transactions. 
Additionally, these information extraction processes can be made more streamlined using an off-chain read-execute-transact-erase-loop~\cite{shukla2021blockiot-retel}, which will ensure that only the necessary data is processed and stored on-chain, optimizing resource utilization and maintaining privacy, while being up-to-date and providing higher throughput.

\section{Related Work}

Emerging technologies and innovative approaches also significantly enhance the security and efficiency of cross-border payments. For instance, Vinayak et al. \cite{vinayak2019design} propose a solution for the automation and transparency of smart contracts on distributed ledgers, similar to our work. However, our research offers a more comprehensive and globally applicable solution compared to their narrower focus on collateral management within financial institutions. Furthermore, Narendra and Aghila \cite{narendra2021fortis} focus on a quantum-resistant smart contract model, suggesting a frontier for future-proof financial transactions prioritizing security. However, their model does not address interoperability, a key aspect of our framework designed to ensure seamless global transactions.
Flynn et al.~\cite{flynn2023enabling} presents a data- and language-agnostic system designed to integrate and analyze data from various sources in DeFi with interoperability and scalability in mind, where they showcase how different programming languages can interact with a unified backend system. However, unlike their work, we have focused on international standards such as ISO20022 in achieving interoperability.
Li et al. emphasize using standardized ontological concepts to ensure interoperability and trustworthy data sharing across institutions and mining transaction logs to detect anomalies and misuse~\cite{li2019leveraging}. In our framework for cross-border payments, we aim to achieve similar interoperability and compliance by leveraging international standards such as ISO20022 and blockchain-based transactions.

Additionally, there are several efforts to automatically generate smart contracts, which simplifies the implementation process. For example, Choudhury et al. \cite{choudhury2018auto}, Tateishi et al. \cite{tateishi2019automatic}, and Frantz and Nowostawski~\cite{frantz2016institutions} have all contributed to this field, each focusing on distinct applications and methods. Yet, all share the common objective of translating business logic into executable code. Another such work is the methodology to automate construction payments by formalizing them into smart contracts proposed by Luo, Han, et al.~\cite{luo2019construction}.
However, while these automation approaches facilitate smart contract creation, they lack the precision needed for sensitive tasks like cross-border payments. Given the complexity and critical importance of financial standards like ISO20022 in enabling cross-border interoperable financial transactions, our framework's tailored approach to ISO20022 translations ensures a higher level of expressivity and compliance with regulatory bodies.
Furthermore, Kang et al.~\cite{kang2024using} investigate the potential of Large Language Models (LLMs) to automate health insurance processes by generating smart contracts from textual policies. 
Van Woensel et al.~\cite{van2023translating} explore translating high-level decision logic from clinical practice guidelines into executable smart contracts on blockchain platforms like Ethereum and Hyperledger Fabric. 
Both these works emphasize transforming domain-specific rules and guidelines from healthcare into smart contracts, similar to our goal of implementing cross-border payments and reporting transactions in the finance domain.

\section{Conclusion}

Our proposed blockchain-based framework offers an innovative solution for enhancing the efficiency and security of cross-chain transactions. By integrating smart contracts and web clients, we align closely with the CBPR+ initiative and employ the ISO20022 messaging format to guarantee a uniform user experience globally. Our framework facilitates a seamless, secure, and efficient mechanism for managing cross-chain transactions, which are crucial in today's globalized blockchain economy.

The detailed system architecture and security protocols we have developed ensure that our framework replicates and augments the current CBPR+ process. Through the strategic use of smart contracts on an EVM-compatible blockchain, we enforce stringent access controls and transaction integrity while enabling real-time transactional updates via an advanced event-driven architecture. This process results in a robust system that reduces risks and enhances transparency throughout the transaction lifecycle. Additionally, the flexibility embedded within our smart contract design allows for custom adaptations to meet diverse regulatory and institutional needs, making our system highly adaptable to future changes in the global financial landscape.

Our research contributes to the evolving financial technology landscape by providing a scalable, secure, and efficient framework for cross-chain transactions. 
The adoption of this framework by financial institutions and regulatory bodies can lead to significant improvements in cross-border payment processes, reducing transaction times and costs while enhancing compliance and security. Furthermore, by demonstrating the feasibility and benefits of a blockchain-based approach, we pave the way for future innovations that can build on our foundational work.

Future research could explore the integration of advanced machine-learning techniques for real-time fraud detection and anomaly monitoring within the transaction process. Additionally, expanding the framework to support multiple non-EVM blockchain platforms could further enhance interoperability and user adoption. Collaborations with regulatory bodies to refine and standardize smart contract protocols will be essential to ensure the widespread acceptance and reliability of this technology.

In summary, our blockchain-based framework offers a flexible, customizable, and secure solution for cross-border payments, addressing the inefficiencies and complexities of traditional financial (TradFi) systems. By integrating smart contracts, web clients, and advanced data processing techniques, our framework sets a new standard for global financial transactions, promoting transparency, compliance, and efficiency. This innovation bridges the gap between TradFi and DeFi, leveraging the strengths of blockchain technology to enhance financial operations in both domains. 
By providing a robust, adaptable, and secure solution, we contribute to the broader goal of creating a more transparent and efficient global financial system, which will ultimately benefit consumers, financial institutions, and regulatory authorities alike.
By adopting our framework, financial institutions can benefit from the robustness and transparency of DeFi while maintaining the regulatory compliance and operational stability essential to TradFi, ultimately resulting in a more integrated and efficient global financial ecosystem.


\section*{Resources}
The source code of our implementation is available at \url{https://github.com/blockchain-interoperability/bi-in-finance}, along with a video demonstration of an example transaction using our implemented framework.

\section*{Acknowledgments}
The authors acknowledge the support from NSF IUCRC CRAFT Center research grant (CRAFT Grant \#22008) for this research. The opinions expressed in this publication do not necessarily represent the views of NSF IUCRC CRAFT.
We are also grateful for the advice from our CRAFT Industry Advisory Board members in shaping this work, especially the input from Jack Pouderoyen and Giri Krishnapillai from SWIFT.

\bigskip

\bibliographystyle{IEEEtran}
\bibliography{references}

\end{document}